\documentclass[manuscript]{aastex}	
\usepackage{epsfig}			
\usepackage{graphicx,color}		
\usepackage{amssymb}			
\usepackage{color}			
\usepackage{url}			
\usepackage{amsmath}			
\usepackage{rotating}			
\usepackage{float}			
\usepackage{textcomp}			
\usepackage{dcolumn}
\usepackage{times}
\usepackage{tabularx}
\usepackage[bottom]{footmisc}
\usepackage{tablefootnote}
\usepackage[colorlinks=true,citecolor=blue]{hyperref}
\usepackage[english]{babel}

\begin{document}
\title{Signature of Extended Solar Cycles as Detected from C\MakeLowercase{a}~{\sc ii}~K Synoptic Maps of Kodaikanal and Mount Wilson Observatory}

\author{Subhamoy Chatterjee$^{1}$,
Dipankar Banerjee$^{1,2}$,
Scott W. McIntosh$^{3}$,
Robert J. Leamon$^{4}$,
Mausumi Dikpati$^{3}$,
Abhishek K. Srivastava$^{5}$,
Luca Bertello$^{6}$
}
  \affil{$^{1}$Indian Institute of Astrophysics, Koramangala, Bangalore 560034, India. e-mail: {\color{blue}{dipu@iiap.res.in}}\\
$^{2}$Center of Excellence in Space Sciences India, IISER Kolkata, Mohanpur 741246, West Bengal, India  \\
$^{3}$High Altitude Observatory, National Center for Atmospheric Research, PO Box
3000, Boulder, Colorado 80307, USA\\
$^{4}$Department of Astronomy, University of Maryland College Park, Maryland 20742,
USA\\
$^{5}$Department of Physics, Indian Institute of Technology (BHU), Varanasi-221005,
India\\
$^{6}$National Solar Observatory, 3665 Discovery Drive, Boulder Colorado 80303, USA \\}

\begin{abstract}

In the recent years there has been a resurgence of the study of Extended Solar Cycles (ESCs) through observational proxies mainly in Extreme Ultraviolet. But most of them are limited only to space-based era covering only about two solar cycles. Long-term historical data-sets are worth in examining the consistency of ESCs.
Kodaikanal Solar Observatory (KSO) and Mount Wilson Observatory (MWO) are the two major sources of long-term Ca~{\sc ii}~K digitised spectroheliograms covering the temporal spans 1907-2007 and 1915-1985 respectively. In this study, we detected supergranule boundaries, commonly known as networks, using the Carrington maps from both KSO and MWO datasets. Subsequently we excluded the plage areas to consider only quiet sun (QS) and detected small scale bright features through intensity thresholding over the QS network. 
Latitudinal density of those features, which we named as `Network Bright Elements' (NBEs), could clearly depict the existence of overlapping cycles with equator-ward branches starting at latitude $\approx 55^{\circ}$ and taking about $15\pm1$ years to reach the equator. We performed superposed epoch analysis to depict the similarity of those extended cycles. Knowledge of such equator-ward band interaction, for several cycles, may provide critical constraints on solar dynamo models. 
\end{abstract}


\section{INTRODUCTION}

Solar cycles are representatives of periodic variation of solar magnetic activity. Different features such as sunspots, plages observed in different wavelength bands serve as potential proxies for representing the solar magnetic cycle. The latitudinal locations of such features when plotted against time produce the well-known `Butterfly diagram'. The most prominent patterns in a Butterfly diagram i.e. equator-ward branches repeat over a time scale of 11 years on an average. These patterns generated from sunspot and plages depict negligible temporal overlap between two subsequent cycles. But the scenario changes when one starts observing the smaller scale magnetic features. Several studies in past have presented observations of small scale features such as ephemeral regions \citep{{1988Natur.333..748W},{2010ApJ...717..357T}}, coronal bright points \citep{2014ApJ...792...12M} to depict a temporal overlap of subsequent equator-ward branches lasting more than half-a-decade and named them as `Extended Solar Cycles (ESCs)'. Apart from those small scale features, signature of ESCs was observed by \citet{1997SoPh..170..411A} from latitudinal distribution of coronal green line (Fe XIV) emission. This study was later substantiated by \citep{2013SoPh..282..249T} with more data corresponding to different coronal heights. They also pointed out that the onset of high latitude activity coincides with current cycle maxima. Another study by \citet{1998SoPh..183..201J} presented an evidence for 17-year solar cycle in coronal hole topology and interplanetary magnetic field directions at 1 AU. However there are arguments that ESCs seen from coronal emissions are merely poleward concentration of trailing polarity flux of old cycle and not the precursor of new solar cycle \citep{2010ApJ...716..693R}. Signature of ESCs were also found from residual of solar differential rotation i.e. torsional oscillation or latitudinal migration of subsurface zonal flow bands \citep{{1982SoPh...75..161L},{1987Natur.328..696S},{2009LRSP....6....1H}}. 
Thus, observation of ESCs across wide range of heights indicates an obvious connection with sunspot cycle.
However, the evolution of ESCs to sunspot butterfly diagram is something which has not been studied well with dynamo theory. This study requires linking of different scales of features and may well provide some missing links to the dynamo models. A detailed study on ESCs may therefore allow us to constrain  the solar cycle predictions \citep{2007AN....328.1027T} and to better estimate their impact on space climate.  ESCs, having temporal overlap of two active latitude bands, can play a critical role in substantiating the solar cycle impact on latitudinal origins of CMEs \citep{{2015NatCo...6E6491M},{2007ApJ...667L.101S}}. A uniform historical data in such case can be extremely useful to know the consistency of such temporal overlaps between two cycles. Several observatories such as Kodaikanal, Mount Wilson, Meudon, Arcetri etc. archived choromospheric data for several cycles before the space era. Successful ESC observations through EUV bright points \citep{2014ApJ...792...12M} inspired us to look at the small scale brightening in Ca~{\sc ii}~K networks as deep rooting of EUV bright points could possibly indicate the connection with Ca~{\sc ii}~K networks.
Among all observatories, Kodaikanal Solar Observatory (KSO) and Mount Wilson Observatory (MWO), archived the longest time-span of Ca~{\sc ii}~K images in digitised form.  We used those two data-sets in this study.
 
\section{OBSERVATIONAL DATA DESCRIPTION}
Full disc Ca~{\sc ii}~K spectroheliograms (1907-2007) observed by KSO has recently been digitised and calibrated. \citet{2016ApJ...827...87C} have generated Carrington maps of size 1571 pixels $\times$500 pixels in Carrington longitude versus sin(latitude) grid (Figure~\ref{bpt_det}a) using the calibrated data. A total of 1184 (between rotation number 716 and 2000) KSO carrington maps have been used in this study. To find consistency of our detection and to improve the statistics we also used a total of 928 (from rotation 827 to 1763) MWO Ca~{\sc ii}~K Carrington maps (Figure~\ref{bpt_det}b) between the years 1915 and 1985 \citep{2011ApJ...730...51S}.  

Previous study by \citet{2016ApJ...827...87C} from KSO showed clear degradation of plage area cycle after the year 1990. This is attributed to the poor data quality and data gaps, resulting in low density of data points (see Figure 1 of \citet{2016ApJ...827...87C}). As bright regions on networks are much smaller scale features compared to plages, it is more likely that data artifacts will affect them more. Similarly, \citep{2014ApJ...793L...4P} also restricted their detection of polar network elements from KSO Ca~{\sc ii}~K data till cycle 21. Furthermore, since we have detected the network elements from Carrington maps the data gaps and  discontinuity of data affects detection significantly. For these reasons, we have restricted our analysis from KSO data till 1990 i.e. the Carrington rotation 1837 (starting on December 19, 1990).

\section{ ANALYSIS OF DATA}
In the following subsections we describe the steps of the data analysis and the corresponding results.
\subsection{Detection of QS Network Bright Elements}
\begin{figure*}[!htbp]
\centering
\vspace{-0.05\textwidth}
\includegraphics[scale=0.75]{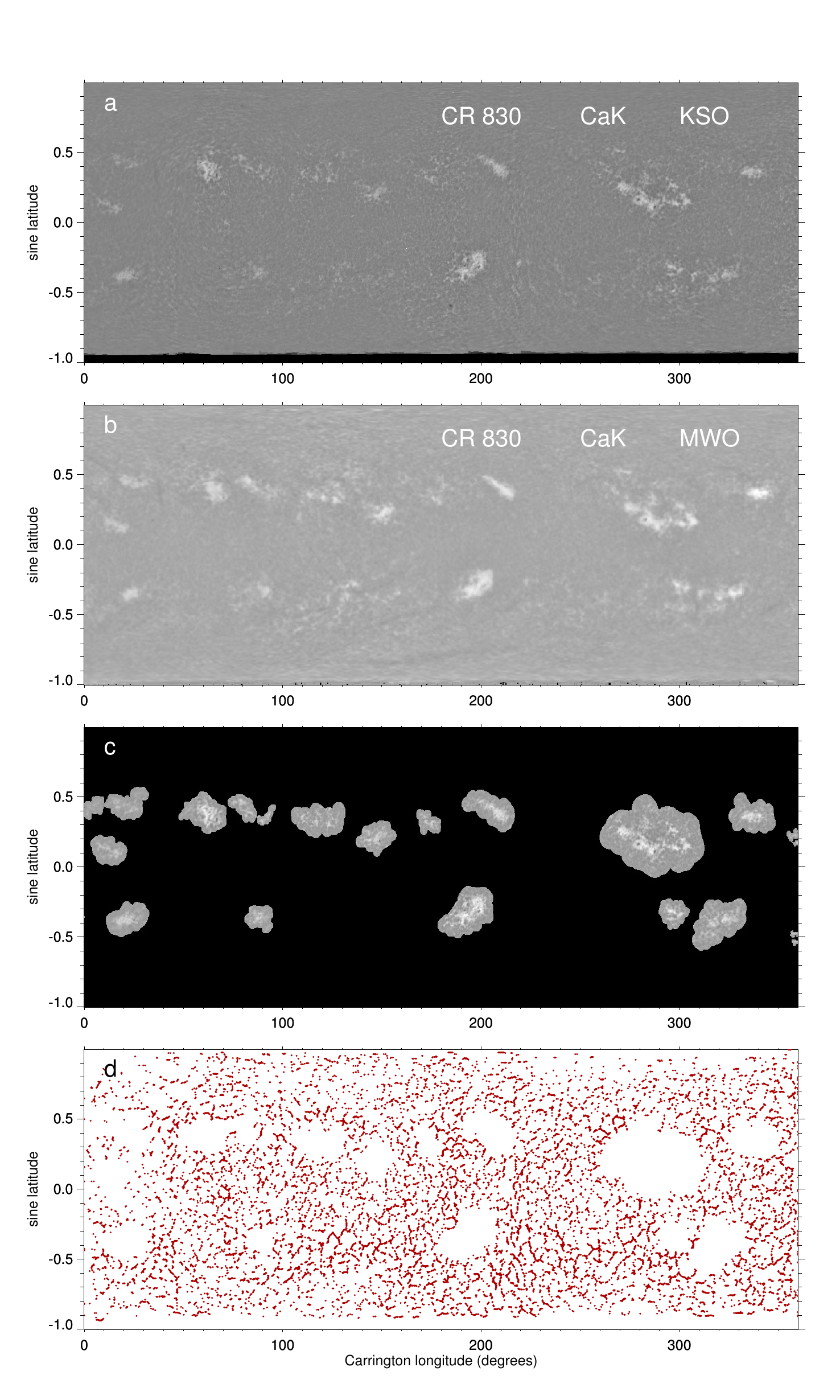}
\vspace{-0.05\textwidth}
  \caption{Observational Data and detection of network bright elements (NBEs) from Carrington maps. (a) Representative Carrington map for rotation 830 generated from Kodaikanal (KSO) Ca~{\sc ii}~K full-disc images; (b) Same Carrington rotation in Ca~{\sc ii}~K from Mt. Wilson Observatory (MWO); (c) Plage areas for rotation 830 masked with some margin outside depending on the size of plage to segregate the QS for NBE detection; (d) Binary map of detected NBEs (red symbols) by thresholding over networks. Please note that symbol size has been chosen for visual representation of NBE locations and is not the accurate representative of the area covered by NBE.}
 \label{bpt_det}
\end{figure*}

Networks manifest as bright polygonal structures in the Ca~{\sc ii}~K images circumscribing the supergranules. The supergranules have been detected from Kodaikanal images by \citet{2017ApJ...841...70C} through watershed transform \citep{2011ApJ...730L...3M}. The watershed transform sees the gray-scale image as a topographic surface with height as intensities. The surface consists of hills and valleys. Watershed transform, also known as `basin finding algorithm', detects the local minimas of the surfaces and fills the valleys unless neighbouring ones start overlapping. Consequently, all the supergranular structures are detected separated by networks. We have used the same technique to detect the networks of the Carrington maps. After extracting the networks, the active regions/plages were blocked with bigger masks. Plages were detected from the Carrington maps using the method described in \citep{2016ApJ...827...87C}. For creating the masks, the plage binary maps were dilated by a circular Kernel having size proportional to the size of the plages (Figure~\ref{bpt_det}c). This was done to avoid any effect of active regions on bright element detection. After blocking the plages, an intensity threshold of (median$+0.5\sigma$) was applied on the Quiet Sun (QS) network to extract the brightenings which we named as `Network Bright Elements (NBEs)' (Figure~\ref{bpt_det}d). The median and standard deviation($\sigma$) were calculated over the QS network intensities.

\subsection{Time-Latitude Diagram of NBE Density}
\begin{figure*}[!htbp]
\hspace*{-2cm}  
\centering
\includegraphics[scale=1.4]{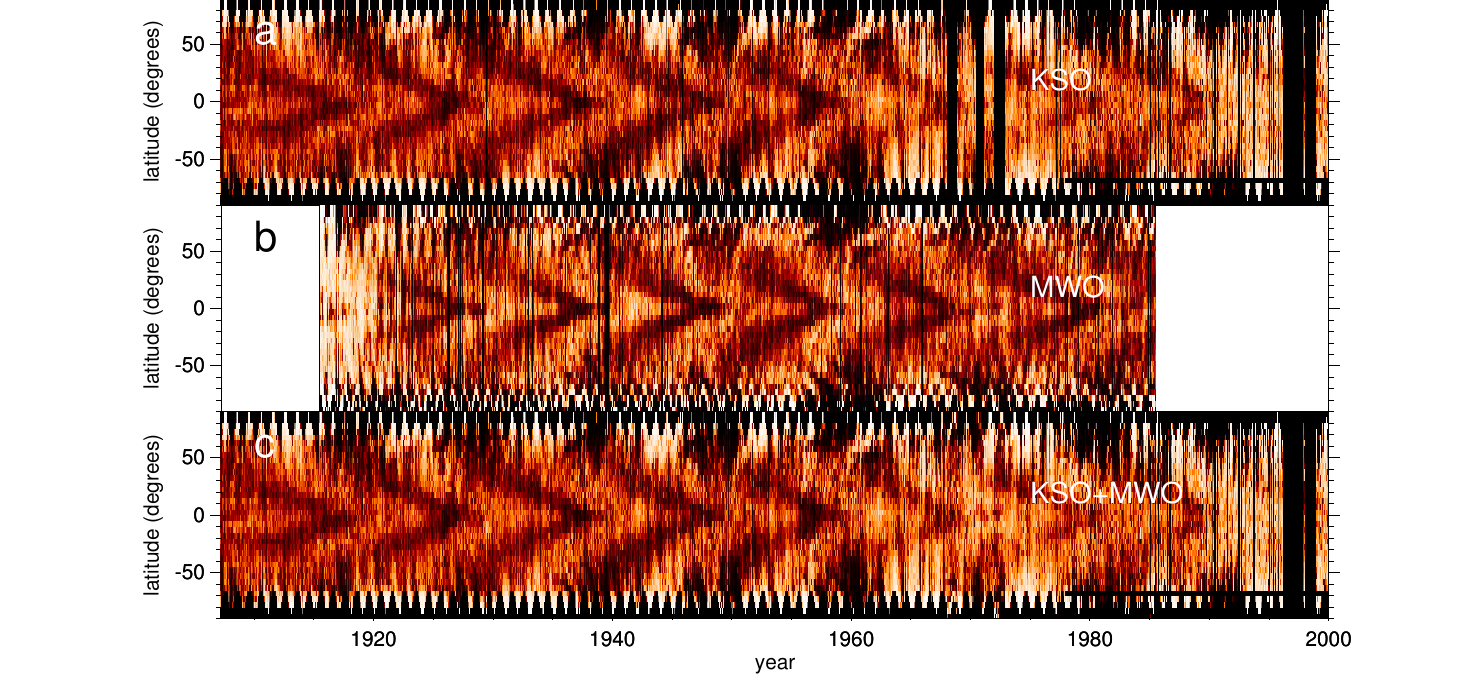}
  \caption{NBE time latitude distribution. (a) Time-latitude plot of NBE density from KSO data; (b)Time-latitude plot of NBE density generated from MWO data; (c) Time-latitude plot of NBE density from KSO with gatagaps after 1960 filled with MWO time-latitude. }
 \label{bpt_time_lati}
\end{figure*}

For generating the time-latitude diagram, first each Carrington map was divided into 36 latitude strips each 5$^\circ$ wide. Then, for $i^{th}$ strip the QS NBE density ($d_i$) was calculated using latitudinal area covered by NBEs ($A_i$), the QS area ($Q_i$) and data gap ($G_i$) as $d_i=\frac{A_i}{Q_i-G_i}$.
This step provided the latitudinal density of QS NBEs as 36 element vector for each Carrington map. These vectors stacked over all the rotations resulted in the time-latitude map of NBE densities. Before, deriving the final map, we smoothed every latitude bin over time with a kernel 200 rotations wide and divided the original map with smoothed map to equalise the contrast across latitudes. The final maps generated from KSO and MWO data are shown in Figure~\ref{bpt_time_lati}a and Figure~\ref{bpt_time_lati}b both clearly depicting extended equator-ward branches having substantial temporal overlap with previous and next. The consistency of ESC detection from KSO and MWO is depicted in Figure~\ref{bpt_time_lati}c. It shows that the KSO extended equator-ward branch for cycle 20 is completed when the gap in KSO NBE time-latitude is filled with MWO NBE density of overlapping times.

\subsection{``Terminators" and Extended Equator-ward Branches}
\begin{figure*}[!htbp]
\hspace*{-2cm}  
\centering
\includegraphics[scale=0.5]{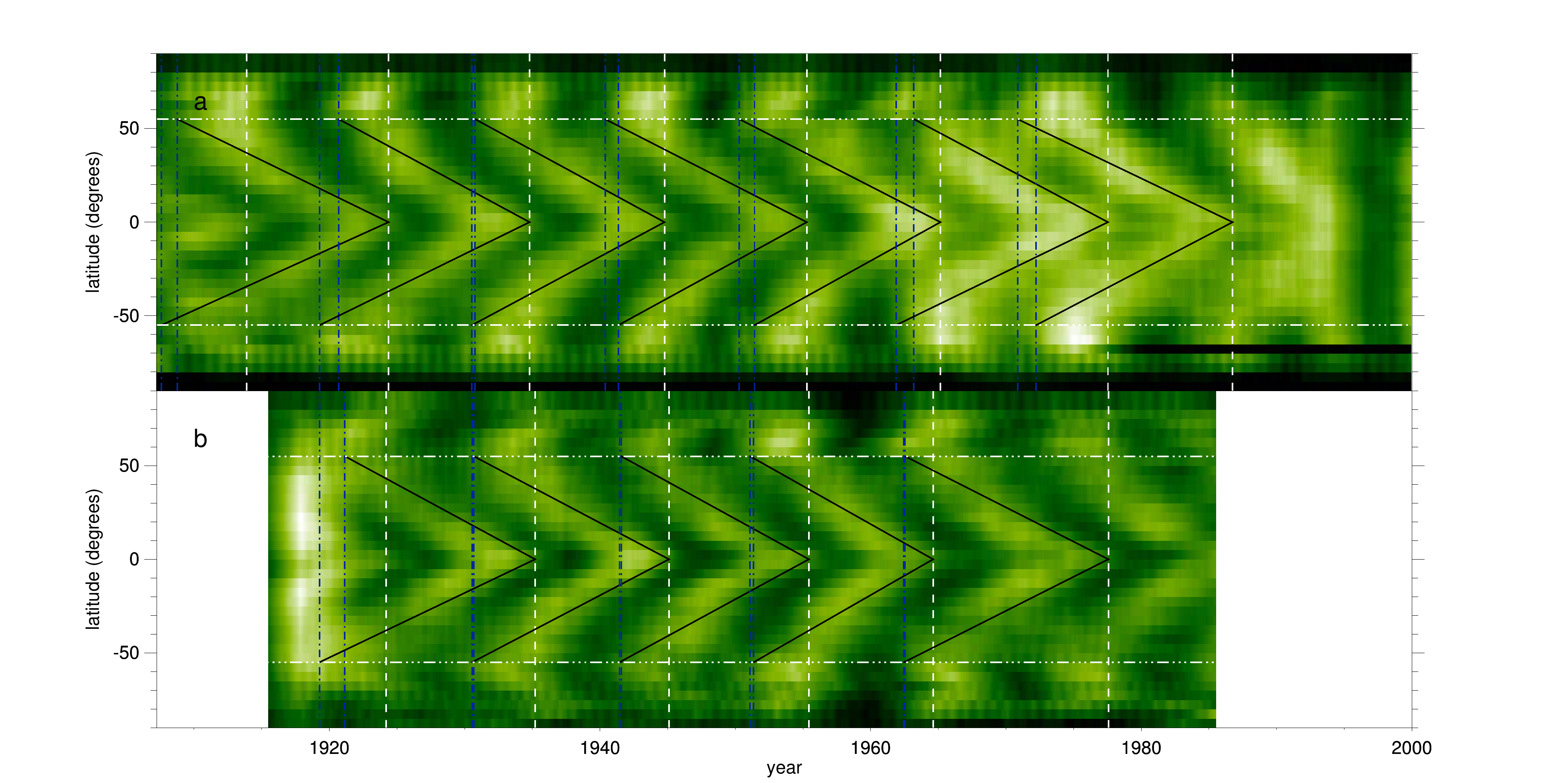}
  \caption{Fitted equator-ward branches over-plotted on smoothed time-latitude plot of NBE density. (a) Fits for KSO NBE time-latitude plot; (b) Fits for MWO NBE time latitude plot. Datagaps in KSO time-latitude is filled with those from MWO before smoothing. Horizontal white dashed lines depict $\pm55^\circ$ latitudes. Vertical blue dashed lines depict the onset of extended equator-ward branches and the white vertical dashed lines represent the epoch of equator-ward branches reaching the equator i.e. ``Terminators". }
 \label{bpt_termin}
\end{figure*}

We defined ``Terminators" as the epochs where the extended cycles come to an end meeting the solar equator \citep{2019arXiv190109083M}. Before detecting ``Terminators", we smoothed the different latitudes of Figure~\ref{bpt_time_lati}b,c with kernel of width 60 rotations to generate the plots shown in Figures~\ref{bpt_termin}a,b. We fitted a third degree polynomial to the smoothed NBE density at the equator within a time span of $[T_{init}-4, T_{init}+4]$ years having form $NBE=p_0+p_1T+p_2T^2+p_3T^3$. Here, $T_{init}$ defines the initial guess about Terminators. We defined the final values of Terminators as the points of inflection of fitted polynomials (Table~\ref{termin_eq_br}) mathematically defined as $T=-p_2/3p_3$. Using the one sigma uncertainties ($\Delta p_2, \Delta p_3$) of the polynomial coefficients from fitting we calculated the uncertainties in terminator locations as $\Delta T=\sqrt{(1/3p_3)^2\Delta p_2^2+(p_2/3p_3^2)^2\Delta p_3^2}$.
The smoothed plots (Figures~\ref{bpt_termin}a,b) especially the one from MWO clearly depicts the bifurcation into a pole-ward branch and an equator-ward branch at $\approx\pm55^\circ$ latitude consistent with study by \citet{{2014ApJ...792...12M},{2019arXiv190109083M}}.  

Thus to deduce the the time-duration of extended equator-ward branches, we also determined the epochs at which those equator-ward start ($S$) at $\pm 55^\circ$ latitude and uncertainties ($\Delta S$) in those for north and south separately using the same method as used for determining ``Terminators". By connecting those start epochs at $\pm 55^\circ$ and the ``Terminators" at equator, we found the fits for equator-ward branches (Figures~\ref{bpt_termin}a, b). The duration of equator-ward branches for different cycles, presented in Table~\ref{termin_eq_br}, were calculated as $D=T-S$. The uncertainties in the cycle duration ($\Delta D$) were derived using quadrature sum of start epoch uncertainty and terminator uncertainties i.e. $\Delta D^2=\Delta S^2+\Delta T^2$. 
It can be observed from both Figure~\ref{bpt_termin}a and Figure~\ref{bpt_termin}b that equator-ward branches of cycle-19 and cycle-20 had the least temporal overlap. Also MWO data shows smaller asymmetry between north and south equator-ward branches with more uniform error bars for all cycle as compared to those for KSO (Table ~\ref{termin_eq_br}). We find good agreement between ``Terminators" determined from KSO and MWO (Table ~\ref{termin_eq_br}, Figure~\ref{bpt_termin}).

\begin{table}
\begin{center}
\caption{Terminator Epochs and Cycle-wise time-duration of equator-ward branches}\label{termin_eq_br}
\resizebox{1.1\columnwidth}{!}{%
\begin{tabular}{lllllll}
\hline
{Cycle} & {Terminators (year)} & &Duration (years) &  &  &  \\ \cline{2-7} 
 & {KSO} & {MWO} & KSO &  & {MWO} \\ \cline{4-7} 
 &  &  & North & South & North & South \\ \hline 
 14 & 1913.88 $\pm$ 0.34 & \\
 	 15 &  1924.38 $\pm$ 0.54 & 1924.19 $\pm$ 0.72 & 15.61 $\pm$ 0.56 & 16.79 $\pm$ 0.55 &&\\
	 16 &  1934.80 $\pm$ 0.91 & 1935.21 $\pm$ 0.74& 14.11 $\pm$ 1.11 & 15.52 $\pm$ 1.30 & 14.06 $\pm$ 1.95 & 15.94 $\pm$ 1.88 \\
	   17 & 1944.77 $\pm$ 1.47 & 1945.10 $\pm$ 0.66& 14 $\pm$ 1.72 & 14.23 $\pm$ 1.63 & 14.43 $\pm$ 1.08 & 14.55 $\pm$ 1\\
	  18 &  1955.28 $\pm$ 1.17 &1955.44 $\pm$ 1.80 &  14.9 $\pm$ 1.65 & 13.92 $\pm$ 1.77 & 13.87 $\pm$ 2.11 & 14 $\pm$ 2.28\\
	19 &  1965.16 $\pm$ 2.96 &1964.62 $\pm$ 2.12 & 14.86 $\pm$ 3.91& 13.74 $\pm$ 3.23 & 13.53 $\pm$ 2.37 & 13.27 $\pm$ 2.53\\
	 20   &  1977.54 $\pm$ 3.57 & 1977.57 $\pm$ 1.93 &  14.35 $\pm$ 5.02 & 15.66 $\pm$ 4.04& 15.03 $\pm$ 2.63 & 15.11 $\pm$ 2.70\\
	  21 &  1986.75 $\pm$ 3.73& &   15.89 $\pm$ 4.76 & 14.52 $\pm$ 4.58 &&\\ \hline
	  Mean&&& 14.82 $\pm$ 1.20 & 14.91 $\pm$ 1.06 & 14.18 $\pm$ 0.94 & 14.57 $\pm$ 0.97 \\\hline
\end{tabular}
}
\end{center}
\end{table}

\subsection{Similarity of Overlapping Cycles: ``Superposed Epoch Analysis"}
\begin{figure*}[!htbp]
\hspace*{-2cm}  
\centering
\includegraphics[scale=0.75]{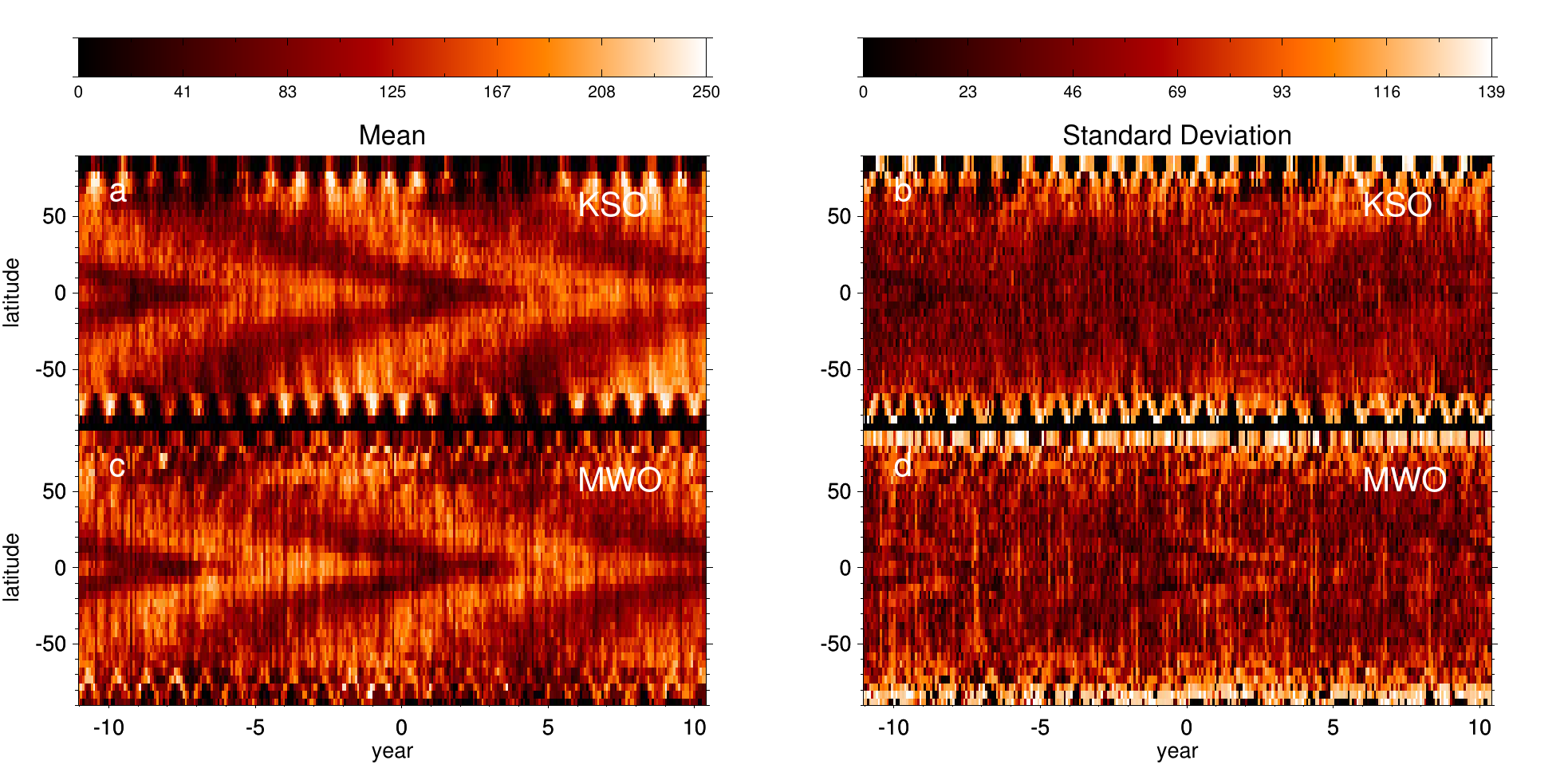}
  \caption{Superposed Epoch Analysis on NBE time-latitude distribution. (a) Mean of 5 epochs defined by $\pm11$ years about terminators for KSO; (b) Standard deviation of 5 epochs for KSO; (c) Mean of 4 epochs defined by $\pm11$ years about terminators for MWO; (d) Standard deviation of 4 epochs for MWO. Figure Color scales are put above the top panels tracing the range of statistical moments.}
 \label{bpt_sea}
\end{figure*}

Superposed Epoch Analysis (SEA) is used to reveal periodicity in a data series through evaluation of statistical moments and also helps to recover signal from noise \citep{1913RSPTA.212...75C}. As the name suggests, this method works by superposition of several epochs of equal length selected about a reference time called `key time'. To select the basis for SEA in our study we selected ``Terminators" as the `key times'. Epochs were selected as the time-latitude distributions within the interval [Terminator$-$11 years, Terminator$+$11 years]. 5 such consecutive epochs were selected starting from cycle-15 Terminator for KSO and 4 consecutive epochs were selected for MWO starting from cycle-16 Terminator. Subsequently, the epochs were averaged to depict the mean epoch (Figures~\ref{bpt_sea}a,c). Mean epoch for both KSO and MWO clearly depicted the significant temporal overlap of successive equator-ward branches confirming the consistency of ESCs over several cycles. The standard deviation over 5 epochs for KSO and 4 epochs for MWO were calculated to depict the dissimilarity of the cycles (Figures~\ref{bpt_sea}b,d). Within the equator-ward branches there is no noticeable variation of standard deviation for both KSO and MWO. 

\section{SUMMARY AND CONCLUSION}

The salient features can be summarised as:
\begin{itemize}
    \item Carrington maps from KSO and MWO were processed to automatically detect NBEs for about 8 and 6 solar cycles respectively.
    \item Latitudinal density of QS NBEs clearly depicted signature of ESCs illustrating the coexistence of two equator-ward branches for the cycles. Also the time-latitude distribution revealed close match between MWO and KSO.
    \item A latitude $\approx 55 ^\circ$ was observed to be the separator between poleward branch and equator-ward branch.
    \item Parameters such as ``Terminators" and time duration of equator branches were extracted from the time-latitude distribution of NBEs from both KSO and MWO. Those parameters showed close match within error bars for both the observatories. The equator-ward branches were seen to originate at $\pm 55^\circ$ latitudes every 11 year taking more than $\approx 14\pm 1$ years to reach the equator.
    \item Superposed Epoch Analysis with ``Terminators" as key times clearly illustrated the similarity of the ESCs with a considerable match between MWO and KSO.
    \end{itemize}
We should point out that Cycle-20 on-wards, KSO showed higher error bars both in terminator epochs and cycle duration as compared to those from MWO. This might be the effect of poor data density and quality of KSO from cycle-20  that reduced the contrast between equator-ward branches after 1965. It should be noted that images of KSO and MWO are not cross-calibrated. For that very reason we treated them separately in this study. Cross-calibration will allow us to fill the gaps in individual Carrington maps and generate even more reliable and consistent results. We wish to pursue that as a future work. For both MWO and KSO least temporal overlap is seen between equator-ward branches of cycle-19 and 20 (Figure~\ref{bpt_termin}). In contrast, cycle-19 is observed as the strongest cycle in last century. Also, previous studies on extended cycles depict a short time difference between sunspot cycle maxima and onset of extended equator-ward branch \citep{2014ApJ...792...12M}. This gives an indication that magnetic interaction between the two equator-ward bands might modulate sunspot cycle amplitude to certain extent based on the duration of temporal overlap and may also contribute in the decay rate of sunspot cycle.

Thus our findings from the Ca~{\sc ii}~K historical data-sets are on par with the results on ESCs from space based data and coronal green line observations. We could validate those results for several cycles starting from cycle 15. Thus, this study established the fact that NBEs detected from chromopsheric networks are effective proxies for ESCs. Such consistent observation of overlapping cycle may provide valuable inputs to traditional dynamo models. Coexistence of two equator-ward branches and their magnetic interaction can play a role in better prediction of sunspot cycles as well space weather with possible impact on Earth's climate \citep{{2015NatCo...6E6491M},{dikpati2019triggering}} .

  \section{ACKNOWLEDGEMENTS}
  We would like to thank all of the observers at Kodaikanal for
their contributions to build this enormous resource over the last
100 years.
KSO data is now available
for public use at  http://kso.iiap.res.in/data.  We also thank the
Science \& Engineering Research Board (SERB) for the project
grant (EMR/2014/000626). We would also like to acknowledge the IUSSTF/JC-011/2016 project grant for supporting this work. We are grateful to the team members of the project on Reconstructing Solar and Heliospheric Magnetic Field Evolution Over the Past Century supported by the International Space Science Institute (ISSI), Bern, Switzerland for giving important suggestions to this project.


   \end{document}